\documentclass[runningheads]{llncs}
\usepackage{graphicx}
\usepackage{amsmath,amssymb,amsfonts}

\usepackage{longtable}
\usepackage{cite}
\usepackage{orcidlink}
\usepackage{booktabs}
\usepackage{tabularray}
\usepackage{lineno}
\usepackage{algorithm2e}
\begin{document}
%
\title{Data-Free Model Extraction Attacks in the Context of Object Detection\thanks{This paper represents the combined and equal contributions of Harshit Shah and Aravindhan during their internship at AIShield.. }}
\titlerunning{DFME Attacks on Object Detection}
%
\author{Harshit Shah\inst{1,2}\orcidlink{0000-0001-8075-9655} \and
Aravindhan G\inst{1,3} \orcidlink{0009-0002-5913-5535} \and
Pavan Kulkarni\inst{1}\orcidlink{0000-0002-8458-6795} \and 
Yuvaraj Govidarajulu \inst{1}\orcidlink{0000-0002-4247-4410}\and Manojkumar Parmar\inst{1}}

\authorrunning{H. Shah. et al.}
\institute{AIShield, Bosch Global Software Technologies Pvt. Ltd., Bangalore, India
\email{\{pavan.kulkarni, govindarajulu.yuvaraj, manojkumar.parmar\}@in.bosch.com}\\
\url{https://www.boschaishield.com/} \and
Nirma University, Ahamdabad, India- \and
Coimbatore Institute of Technology, Coimbatore, India\\
\email{\{harshit1409, aravindganpath\}@gmail.com}}
\maketitle              
\begin{abstract}
A significant number of machine learning models are vulnerable to model extraction attacks, which focus on stealing the models by using specially curated queries against the target model. This task is well accomplished by using part of the training data or a surrogate dataset to train a new model that mimics a target model in a white-box environment. In pragmatic situations, however, the target models are trained on private datasets that are inaccessible to the adversary. The data-free model extraction technique replaces this problem when it comes to using queries artificially curated by a generator similar to that used in Generative Adversarial Nets. We propose for the first time, to the best of our knowledge, an adversary black box attack extending to a regression problem for predicting bounding box coordinates in object detection. As part of our study, we found that defining a loss function and using a novel generator setup is one of the key aspects in extracting the target model. We find that the proposed model extraction method achieves significant results by using reasonable queries. The discovery of this object detection vulnerability will support future prospects for securing such models.

\end{abstract}

\section{Introduction}

The advent of artificial intelligence (AI) has revolutionized the world, bringing numerous benefits by automating various tasks. However, these advancements have also created new opportunities for exploitation and risk. AI models are vulnerable to a wide range of attacks, and one such novel method of attack is adversarial AI. Adversarial attacks can manipulate the model, carefully poison input data, or use queries to create a copy of the original model. Adversarial attacks on AI models have evolved significantly over the years, and they can be categorized into three different methods: evasion, poisoning, and extraction. Evasion attacks are designed to fool the AI model into making a wrong decision, while poisoning attacks introduce a small number of malicious inputs into the training data to manipulate the model's decision-making process. Extraction attacks, on the other hand, aim to extract the model's information without accessing the original training data.

Object detection is an essential application of AI, with widespread use in various fields, including surveillance, autonomous vehicles, and robotics. Object detection models can detect and locate objects of interest within an image, making it a crucial component in the development of intelligent systems. However, adversarial attacks on object detection models can have severe consequences. For example, an attacker can manipulate the model to ignore or misidentify specific objects, leading to security breaches or accidents in autonomous systems. Previous research has proposed various methods to evade object detection and image segmentation models \cite{xie2017adversarial}. Other research has proposed using text generation to mimic a model extraction attack on object detection models \cite{liang2022imitated}. However, to the best of our knowledge, no previous research has explored model extraction attacks in object detection tasks in a black-box environment.

The remainder of the paper is organized as follows: Section 2 presents the literature survey. Section 3 presents the concepts of related work, DFME and object detection. Section 4 presents a detailed analysis of our proposed methodology and algorithm. In Section 5, we describe the dataset used and present the experimental results. Finally, Section 6 concludes the paper by discussing its limitations and future research directions.

\section{Literature}

The vulnerability of AI models has been explored by several researchers who have proposed various methods to secure them. Szegedy et al. demonstrated the susceptibility of deep neural networks to adversarial examples, which are perturbed inputs used in three types of attacks: evasion, poisoning, and extraction  \cite{szegedy2013intriguing}. Biggio et al. provided an overview of adversarial machine learning and emphasized the imperceptibility of perturbations used in evasion and poisoning attacks to the human eye \cite{biggio2018wild}. The authors also highlighted the exploitation of Machine Learning as a Service (MLaaS) environments by adversaries. 

Multiple attacks have been proposed to steal machine learning or deep learning models from MLaaS platforms, using a prepared dataset to train a clone model from the predictions obtained from the original model \cite{tramer2016stealing, yu2020cloudleak, chandrasekaran2020exploring, yan2020cache}. Yu et al. proposed a transfer learning-based approach to minimize the queries required for this process \cite{yu2020cloudleak}, while Yan et al. suggested a cache side-channel attack for stealing the architecture of Deep Neural Networks (DNNs) \cite{yan2020cache} . These attacks pose a significant threat to the security of machine learning systems, highlighting the need for effective defense mechanisms to mitigate their impact.

The concept of data-free model extraction (DFME) is an extension of knowledge distillation \cite{chen2021distilling, gou2021knowledge, hinton2015distilling}, where knowledge is transferred from a target model to a stolen model without using any dataset. Hinton et al. proposed distillation, which uses teacher logits and ground truth to monitor student learning \cite{hinton2015distilling}, while Romero et al. and Zagoruyko et al. improved distillation's effectiveness in \cite{romero2014fitnets, zagoruyko2016paying}. Truong et al. introduced DFME, which uses zero-order optimization (ZOO) to extract a trained classification model in a black-box setting \cite{truong2021data}. Kariyapa et al. emphasized the importance of using loss functions like KullbackLeibler divergence and $l_1$ norm loss when using ZOO for model extraction \cite{kariyappa2021maze}. Miura et al. demonstrated that DFME attacks are possible on gradient-based explainable AI, where explanations are used to train the generative model to reduce the number of queries needed to steal the model \cite{miura2021megex}.

In the object detection domain \cite{ge2021yolox, ren2015faster}, where the output of an object detection model differs significantly from a classification model, attacks on object detection and image segmentation models can be challenging. Xie et al. proposed the Dense Adversary Generation (DAG) technique, which generates adversarial examples for semantic segmentation and object detection \cite{xie2017adversarial}. Liang et al. recently proposed an imitation-based model extraction attack on object detection models using dataless model extraction and text-image generation to generate a synthetic dataset \cite{liang2022imitated}. The authors propose using natural scenes and text-image generation to accelerate the generation of a domain-specific synthetic dataset and then train an imitation object detector on this dataset.

In this paper, we introduce a new technique for extracting object detection models in a black box setting. One critical aspect of object detection is the precise labeling of objects and their corresponding bounding box coordinates. Since these coordinates are the outcome of a regression task, any model extraction attack used for the classification task cannot be directly applied to an object detection task. To the best of our knowledge, this is the first time that a model performing a regression task has been successfully extracted in a black box environment. The proposed method is crucial to improving the security of object detection models, which are essential in developing intelligent systems.

\section{Methodology}
\subsection{Data-Free Model Extraction on Object Detection}

\begin{figure}[t]
  \centering
  \includegraphics[height=140pt, width=350pt]{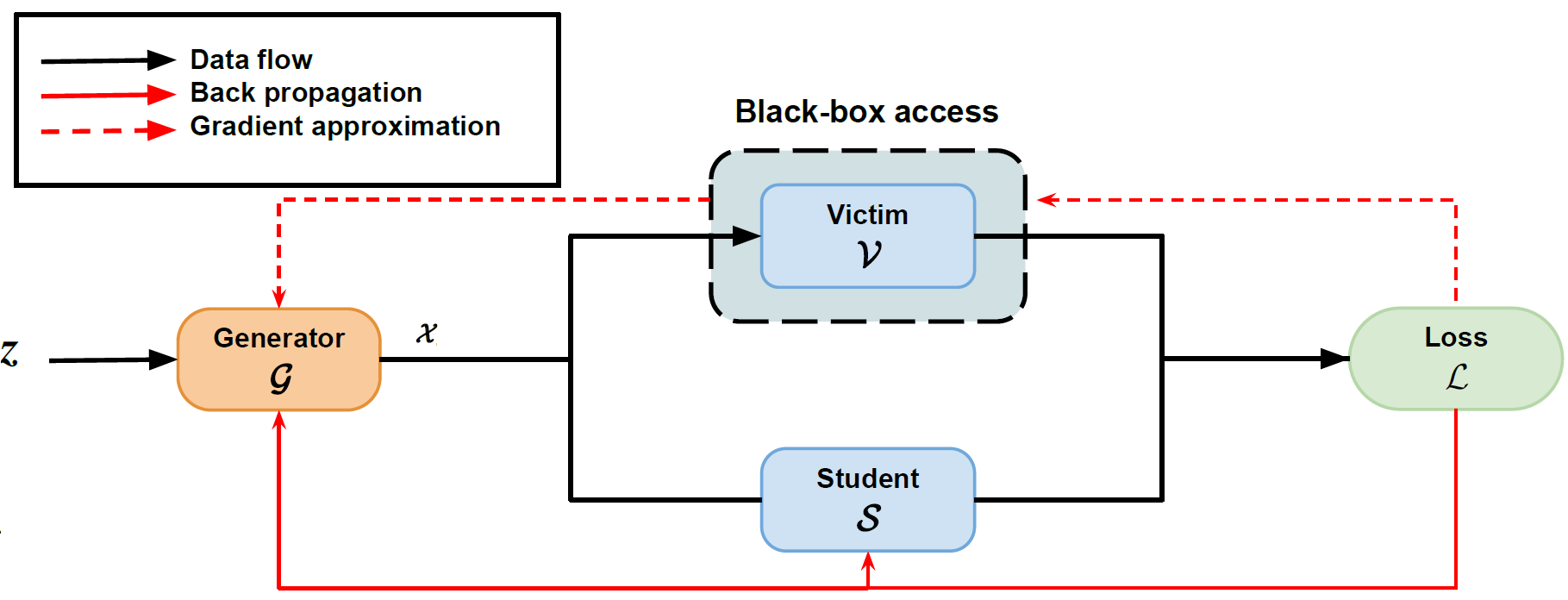}
  \caption{Adapted Data-free Model Extraction Attack\cite{truong2021data} on object detection Framework. }
  \label{fig:Framework}
\end{figure}

The proposed attack setup builds on the DFME attack \cite{truong2021data} and customizes it to extract object detection models. This attack architecture revolves around three key components: a victim model ($V$), a student model ($S$), and a generator ($G$) as shown in Fig. \ref{fig:Framework}. The victim model is a pre-trained model specialized in object detection tasks, while the student model aims to distill knowledge from the victim. The generator, on the other hand, synthesizes data to maximize the error between the victim and student models. By leveraging this architecture, adversarial queries can be crafted to extract essential information from the victim model, even when the attacker has only black-box access to it.

\textbf{Victim Model}: A pre-trained model that is specialized for a task-oriented domain data set $D_v$. Typically, these victim models are accessible through an application programming interfaces (API) call, providing adversaries with only black box access to the model. For object detection tasks, $V$ is trained on a dataset that includes both the class information ($\theta_{cls_i}$) and the bounding box coordinates ($\theta_{bb_i}$) of the objects in the image. The bounding box coordinates are represented as a set of four real values, denoted as
\begin{equation}
    \label{equ:coor}
    \theta_{bb_i} = \left(x_{min_i},y_{min_i}, x_{max_i} ,y_{max_i}\right)
\end{equation}

Therefore, the $V$ predicts two outputs: \textit{label} for classification and \textit{bbox} for regression, which are combined as

\begin{equation}
\begin{split}
\label{equ:function}
\theta_i = {\theta_{cls_i}} + {\theta_{bb_i}}\\
\end{split}
\end{equation}

\textbf{Student Model}:
It is employed to demonstrate knowledge distillation, proving that knowledge from a model can be effectively transferred to another model, even if the latter has a smaller architecture compared to $V$ \cite{ba2014deep, hinton2015distilling}. For this purpose, a pre-trained model is selected as $V$, and another model is chosen as $S$. This setup allows the adversary to choose a suitable student model architecture without needing any prior knowledge of the victim model's architecture, thereby maintaining the black box condition. The results are presented for various student-victim model pairs in later sections. The loss function $l$, based on knowledge distillation, is used to identify disagreement between the predictions of $V$ and $S$. The student model produces three values: \textit{pre-label} (logits before classification activation), \textit{label} (logits after classification activation), and \textit{bbox} (regression coordinates).

\textbf{Generator}:
The traditional architecture of DFME involves a single generative model denoted as $G$. It is responsible for synthesizing data using random vector points ($z$) sampled from a Gaussian (Normal) Distribution ($\mathcal{N}$). The goal of the generator is to maximize the error between $V$ and $S$ by generating synthetic data that highlights their discrepancies. Unlike image classification tasks, where models only need to recognize features, object detection tasks require both semantic and spatial information. Thus, the generator's architecture captures both semantics and spatial information to produce synthetic data that mimics the characteristics of the original dataset. The loss functions used for the generator are the same as those for the student model. However, the generator aims to maximize the error between $V$ and $S$ for predictions made on the synthetic data ($D_s$). This results in a game of minimization and maximization between $G$ and $S$, represented by equations \eqref{equ:minmax} and \eqref{equ:minmax1}.

\begin{equation}
\begin{split}
\label{equ:minmax}
    \underset{S}{min} \underset{G}{max} \ \mathbb{E}_{z} \left[l_{total}(V(G), S(G))\right]
\end{split}
\end{equation}
\begin{equation}
    \begin{split}
    \label{equ:minmax1}
    G =  \mathbb{E}_{z \approx \mathcal{N}(0,1)}
    \\
    \end{split}
\end{equation}

The process worflow is summarized in \figurename \ref{fig:Framework} where the random vector points $z$, sampled from Gaussian and Laplacian distributions, are fed into $G$, which generates a synthetic image $X$. This synthetic image $X$ is then passed as input to both $V$ and $S$ models. Predictions made by $V$ and $S$ models combine probabilistic confidence values of different classes $P_{class_i},\forall class_i\in [0,1]$ with bounding box coordinates of the detected object $B_{box_i}, \forall box_i \in [0,1]$. The classification loss ($l_{cls}$) and regression loss ($l_{reg}$) are calculated based on the predictions of the $V$ and $S$.

The total loss ($l_{total}$) is computed as the sum of the classification and regression losses, as shown in Equation \eqref{equ:total}. During the back-propagation phase, gradients are calculated from the total loss, and they are used to train both $S$ and $G$ by updating their respective weights and biases in each iteration. The aim is to minimize the loss of the $S$ while maximizing the error between $V$ and $S$ on synthetic data.
\begin{equation}
   l_{total} = l_{cls} + l_{reg} 
   \label{equ:total}
\end{equation}

\RestyleAlgo{ruled}
\begin{algorithm}[t]
\IncMargin{1em}
\SetAlgoLined
\DontPrintSemicolon
\SetKwInOut{Input}{Input}
\caption{ DFME on object detection} \label{alg:one}
\Input{Query budget \textit{Q}, generator iteration \textit{$n_{G}$}, student iteration \textit{$n_S$}, learning rate {$\eta$}, latent dimension \textit{d}}
\KwResult{Trained student model \textit{$S$}}

\While{\textit{Q} $> 0$}
{
    \For{$i\leftarrow{1}$ \KwTo \textit{$n_{G}$}}
    {
        $z_{d} \sim \mathcal{N}(0,1)$\\
        $x$ = $G(z_{d};w_{G},b_{G})$\\
       approximate gradient $\nabla_{(w_{G},b_{G})} \ \textit{$l_{total}(x)$}$\\
       $w_{G},b_{G} = ({w_{G},b_{G}}) - \eta \nabla_{(w_{G},b_{G})} l_{total}(x)$

    }
    
    \For{$i\leftarrow{1}$ \KwTo \textit{$n_{S}$}}
    {
       $z_{d} \sim \mathcal{N}(0,1)$\\
       $x$ = $G(z_{d};w_{G},b_{G})$\\
       $X$ = shuffle($x$)\\
       compute $V_{X}$,$S_X$,\textit{$l_{total}(X)$},$\nabla_{(w_{S},b_{S})} \ \textit{$l_{total}(X)$}$\\
       $w_{S},b_{S} = (w_{S},b_{S}) - \eta \nabla_{(w_{S},b_{S})}l_{total}(X)$
    }
    continue remaining query budget \textit{Q}

 }
\end{algorithm}

\textbf{Algorithm.} $G$ and $S$ are trained alternatively at each iteration. To achieve min-max agreement between $G$ and $S$, $G$ is trained on \textit{$n_G$} iterations and $S$ is trained \textit{$n_S$} iterations. In this setting, it is important to avoid over-training of $G$ in order for the $S$ to capture the necessary details effectively. Therefore, the value of generator iteration ($n_G$) is set to 1, while the value of student iteration ($n_S$) is set to 5.
Additionally, several parameters are introduced in the algorithm:
\begin{itemize}
    \item \textit{Q}: Query budget, which represents the number of times data generation is performed in one iteration. 
    \item $\eta$: Learning rate of the networks, which is used to update the parameters as explained in the subsequent sections.
    \item \textit{d}: Latent dimension, indicating the number of random vector points sampled from the distributions.
\end{itemize}

\subsection{Loss functions}
In this section, we provide an explanation of the loss function utilized in our classification and object detection tasks. Prior research on DFME has discussed two approaches for the loss function. The first approach, called distillation, employs the Kullback-Leibler (KL) Divergence represented as $l_{KL}$ and the $l_1$ norm loss, as shown in equation \eqref{equ:kl}, where $N$ denotes the number of classes. It has been noted by Fang \textit{et al.} that when the student learning approaches the victim's values, the KL Divergence loss function encounters the issue of vanishing gradients \cite{fang2019data}.

Additionally, the authors of DFME incorporate the forward differences method as presented in \cite{truong2021data}. Our work extends the distillation method by defining the classification loss $l_{cls}$ using the $l_{1}$ norm loss function, and for object detection, we employ the root mean squared error ($MSE$).


\begin{equation}
\label{equ:kl}
    l_{KL} = \sum\limits_{i=1}^{N} V_i(x)log\left( \frac{V_i(x)}{S_i(x)} \right)
\end{equation}
\begin{equation}
\label{equ:l1}
    l_{1} = \sum\limits_{i=1}^{N} \left| V_i(x) - S_i(x) \right|
\end{equation}

\begin{equation}
\label{equ:mse}
    l_{mse} = \sum\limits_{i=1}^{K} \left| V_i(x) - S_i(x) \right|^2
\end{equation}

\begin{equation}
\label{equ:cross}
    l_{crossentropy} = -\sum_{c=1}^MS_i(x)\log(V_i(x))
\end{equation}

The total loss for object detection is considered as the sum of these two losses $l_{total}$ 
\begin{equation}
\label{eq:total}
    l_{total} = l_{1} + l_{mse}
\end{equation}

\section{Experimental Analysis}
In this section, we discuss the dataset and its preprocessing required, along with the setting of experiments performed. In addition to this, we showcase the final results with evaluation metrics.

\subsection{Dataset Description}
We showcase results for the \textbf{Caltech-101} \cite{1384978} and the \textbf{ Oxford-IIIT Pet Dataset} \cite{parkhi2012cats}. 
There were 101 item categories for Caltech-101, however for the experiment, we selected a sample of 1000 images from the original dataset that belonged to 10 classes. Faces, Leopards, aeroplanes, butterflies, cameras, dalmatians, pizza, revolvers, umbrellas, and wheelchairs were among the labels that were given some thought for the experiment. The Pets dataset consisted of 37 classes of different breeds of animals. To reduce the complexity we reduce the classification to a binary classification with one class as dog and other as cat.
The datasets ground truth
consisted of the class name of the image $D_{V_i}$ and four coordinates corresponding to the object’s
location. These coordinates were $x_{min}, y_{min}, x_{max}, y_{max}$ along
with width W and height H of the image.

\subsection{Preprocessing}
Initially, the Caltech-101 and Oxford Pets images were in varying shapes, necessitating standardization of the image shapes. Subsequently, pixel values were scaled to the range [0,1].

Furthermore, each starting coordinate value $x_{min_i}$ and $y_{min_i}$ were divided by their respective width $W_i$ and height $H_i$, and each ending coordinate value $x_{max_i}$ and $y_{max_i}$ were divided by their corresponding width $W_i$ and height $H_i$.

\begin{equation}
    \begin{split}
    \label{equ:bscale}
        x_{min_i} = \frac{x_{min_i}}{W_i} \hspace{1in}
        x_{max_i} = \frac{x_{max_i}}{W_i}\\
        y_{min_i} = \frac{{min_i}}{H_i} \hspace{1in}
        y_{max_i} = \frac{y_{max_i}}{H_i}\\
    \end{split}
\end{equation}

\begin{table}[t]
\caption{DFME results on Oxford Pets and Caltech101 datasets} 
\centering
\resizebox{\linewidth}{!}{%
\begin{tblr}{
  cell{1}{3} = {c=6}{},
  cell{2}{1} = {c=2}{},
  cell{2}{3} = {c=2}{},
  cell{2}{6} = {c=2}{},
  cell{7}{1} = {c=8}{},
  cell{8}{3} = {c=6}{},
  cell{9}{1} = {c=2}{},
  cell{9}{3} = {c=2}{},
  cell{9}{6} = {c=2}{},
  hline{1-3,7-10,12} = {-}{},
}
Dataset        & \textbf{Oxford Pets} &              &        &                  &          &        &                  \\
Trained Models &                      & Accuracy(\%) &        & Success Rate(\%) & IoU(\%)  &        & Success Rate(\%) \\
Victim         & Student              & Baseline     & Attack &                  & Baseline & Attack &                  \\
VGG16          & VGG16                & 99           & 70     & 70               & 94       & 66     & 70               \\
Resnet50       & VGG16                & 91           & 73     & 80               & 71       & 68     & 95               \\
Inception V3   & InceptionV3          & 99           & \textbf{90}     & 90               & 90       & \textbf{68}     & 75               \\
               &                      &              &        &                  &          &        &                  \\
Dataset        & \textbf{Caltech101}  &              &        &                  &          &        &                  \\
Trained Models &                      & Accuracy(\%) &        & Success Rate(\%) & IoU(\%)  &        & Success Rate(\%) \\
Victim         & Student              & Baseline     & Attack &                  & Baseline & Attack &                  \\
VGG16          & VGG11                & 93           & \textbf{91}     & 98               & 88       & \textbf{82}     & 93               
\end{tblr}
}
\label{table:results}
\end{table}

\subsection{Experimental Setting}

To conduct our experiments, we utilize open-source pre-trained models, namely ResNet-50 \cite{he2016deep}, VGG-16 \cite{simonyan2014very}, and InceptionV3 \cite{szegedy2016rethinking}. From these models, we carefully select the architecture for both $V$ and $S$ models.

For the classification task in the victim model, we employ the widely used cross-entropy loss as defined in Equation (\ref{equ:cross}). On the other hand, for the object detection task, we use the root mean squared logarithmic error (RMSLE) as the loss function. To optimize our models, we choose the adaptive momentum estimator, Adam, as the optimizer. The initial learning rate ($\eta_{V}$) is set to 0.001 and is reduced by a factor of 0.5 after three epochs. The learning rate decay continues until a minimum value of 0.0001 is reached.

To activate the victim's label, we employ the sigmoid function ($\sigma$) as the chosen activation function. Likewise, for both the classification branch and the bounding box task, we use the $\sigma$ as the activation function. This choice is motivated by the need for binary classification, achieved through  $\sigma_{cls}$ for the classification branch. Similarly, for the bounding box values, we require them to be within the range of zero to one, which is achieved by using the sigmoid activation function denoted as $\sigma_{bb}$, as shown in equation \eqref{equ:activation}.

\begin{equation}
\begin{split}
\label{equ:activation}
\sigma(z) = \frac{1} {1 + e^{-V_{y_i}}} \hspace{1in}
y_{i} = G(x) \forall{x} \in{D_v}
\end{split}
\end{equation}

In the context of data-free model extraction for object detection, we refer to Algorithm 1. In this algorithm, the latent dimension (\textit{d}) for random vector points ($z$) is configured to be the same as the batch size, which is 256. These random vector points are utilized to generate synthetic data ($D_S$), resulting in a total of 256 images for each iteration. The value of \textit{Q} is set to 5,000,000, and the number of generator iteration (\textit{$n_G$}) is set to 1, while the number of student iterator (\textit{$n_S$}) is set to 5.

Both $S$ and $G$ are optimized using the Adam optimizer, with the learning rates ( $\eta_{S, G}$) decaying exponentially at rates of 0.8 and 0.96, respectively, for 1000 steps. Through our experiments, we determined that the initial values of $\eta_{S, G}$ should be within the range of [0.02 - 0.0002]. Furthermore, our findings indicate that for improved results, setting the learning rate of the generator ($\eta_G$) higher than the learning rate of the student model ($\eta_S$) helps achieve min-max disagreement. The loss weights for both $V$ and $S$ in the classification and object detection parts are assigned equally. Each attack is executed for a total of 53 iterations, corresponding to a query complexity of 5 million.

\subsection{Evaluation metrics}
In order to evaluate the performance and effectiveness of the attack, we use metrics Accuracy (${y}_{acc}$) \text for assessing both $V$ and $S$ \cite{truong2021data}. 

For the bounding-box regression problem, the IoU (${y}_{iou}$) metric is used \cite{metric_evaluation, iou_accuracy}. It compares the overlap between the predicted bounding box and the ground truth box. 

\begin{figure*}[t]
  \centering
  \includegraphics[height=190pt, width=310pt]{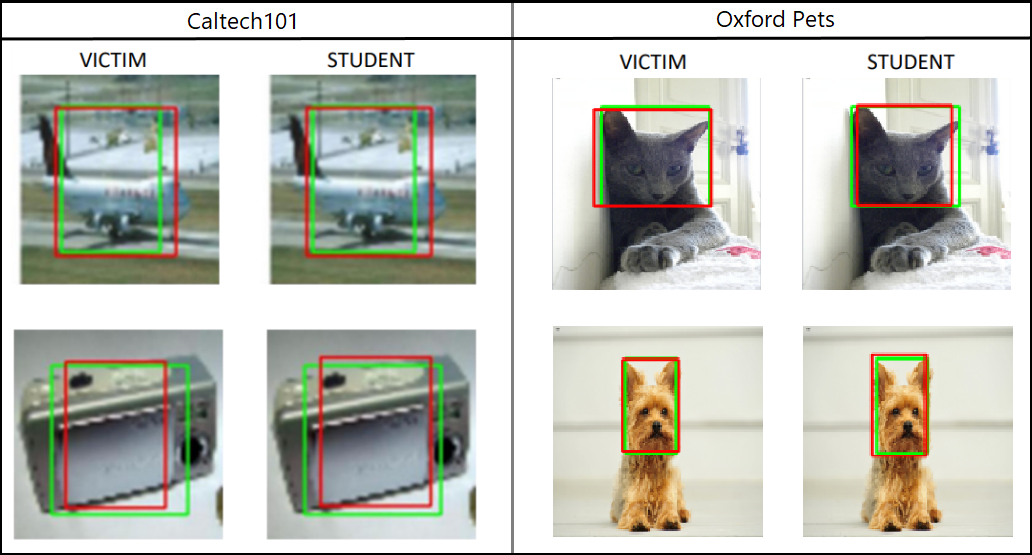}
  \caption{Comparison of victim and student performance with ground truth on Caltech101 and Oxford Pets datasets. The green box represents the ground truth, and the red box represents the prediction by the victim and student models.}
  \label{fig:results_comp}
\end{figure*}

\subsection{Experimental results}

Table \ref{table:results} shows the comparative performance of student models against the victim models. We experimented with different pretrained backbone models. The Baseline evaluation metrics refer to the performance of victim model. It is named Baseline as we compare our student model to the performance of vicitm model. Consequently, the evaluation metrics for student is known as Attack. Success Rate percentage is the efficacy of attack, i.e., the performance of student model (Attack Accuracy) to the victim's accuracy (Baseline Accuracy).
\begin{equation}
\label{equ:success_rate}
    Success Rate = Attack(\%) / Baseline(\%)
\end{equation}
We use the IoU threshold of 0.5 in our experiments \cite{yolov3}. For Oxford Pets dataset, we observe that using a similar backbone model architecture of Inception V3 specifically produces a student model with accuracy of 90\% and relative accuracy (between Student and Victim) of 90\%. The IoU value for the same student model is 68\% and a relative IoU (between Student and Victim) of 75\%. For Caltech 101, the victim model, with VGG16 as a backbone architecture, was able to achieve a classification accuracy of 93\% and 88\% IoU. In comparison to the victim model, the student model underwent the DFME algorithm and did not have access to the original data at all. The student model can attain a classification accuracy of 91\% and IoU of 82\%.

\begin{figure}[t]
  \centering
  \includegraphics[height=120pt, width=170pt]{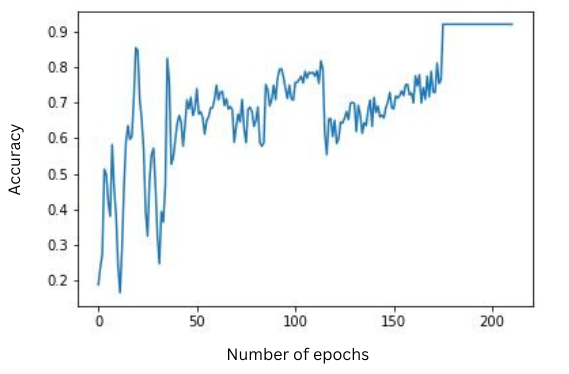}
   \includegraphics[height=120pt, width=170pt]{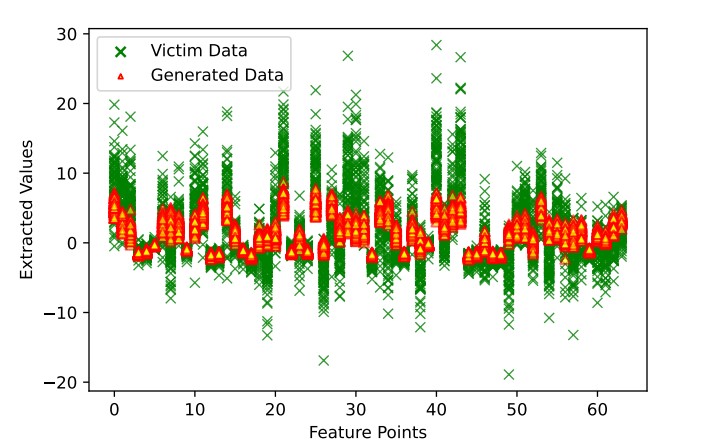} \\
   {(a)} \hspace{2.3in} {(b)}
  \caption{ (a) Accuracy curve of Student model on Caltech101 dataset. (b) Scatter plot on the generated images (in red) along with the original images (in green) }
  \label{fig:curve}
  
\end{figure}

Sample example outputs from the Caltech 101 and Oxford Pets datasets are shown in \figurename \ref{fig:results_comp}. The student model is able to make predictions of bounding box coordinates, that is, the regression branch of object detection as well as the victim model. 
Fig. \ref{fig:curve}(a) depicts the accuracy trend of the student model, illustrating its continuous improvement over multiple epochs as it successfully emulates the victim model. On the other hand, Fig. \ref{fig:curve}(b) showcases the presence of generated images within the victim domain. This observation validates the generator's capability to identify the victim domain and generate images that serve as attack vectors for model extraction.

The primary objective of our research paper is to demonstrate the process of model extraction. Specifically, we aim to highlight the potential vulnerabilities of object detection models to model extraction attacks. The experimentation setup allows for the possibility of expanding to multiple objects if the feasibility for a single object is established. The models selected for this study will serve as a fundamental reference point. Additionally, the scope of the research can be broadened to include transformer-based techniques by altering the student and victim models while maintaining consistency with the generator model.

\section{Conclusion and Future Scope}
We proposed a method based on the DFME technique for stealing object detection models.  By conducting experiments we have shown the feasibility of our approach and revealed existing vulnerabilities in this task. One potential avenue for future exploration is the extension of our method to encompass multiple object detection tasks. This would involve applying the DFME technique to a broader range of datasets and model architectures, providing substantial evidence of the attack's efficacy and generalizability.

Additionally, given the significant implications of model extraction attacks, it becomes crucial to focus on developing robust defense mechanisms. Building defense models that can effectively counter DFME attacks and enhance the security of object detection systems represents a promising direction for future research. By investigating and implementing countermeasures, we can work towards strengthening the integrity and confidentiality of object detection models, ultimately ensuring the privacy and trustworthiness of such systems.

\bibliographystyle{splncs04}
\bibliography{mybibliography}

\begin{thebibliography}{10}
\providecommand{\url}[1]{\texttt{#1}}
\providecommand{\urlprefix}{URL }
\providecommand{\doi}[1]{https://doi.org/#1}

\bibitem{ba2014deep}
Ba, J., Caruana, R.: Do deep nets really need to be deep? Advances in neural
  information processing systems  \textbf{27} (2014)

\bibitem{biggio2018wild}
Biggio, B., Roli, F.: Wild patterns: Ten years after the rise of adversarial
  machine learning. Pattern Recognition  \textbf{84},  317--331 (2018)

\bibitem{chandrasekaran2020exploring}
Chandrasekaran, V., Chaudhuri, K., Giacomelli, I., Jha, S., Yan, S.: Exploring
  connections between active learning and model extraction. In: 29th USENIX
  Security Symposium (USENIX Security 20). pp. 1309--1326 (2020)

\bibitem{chen2021distilling}
Chen, P., Liu, S., Zhao, H., Jia, J.: Distilling knowledge via knowledge
  review. In: Proceedings of the IEEE/CVF Conference on Computer Vision and
  Pattern Recognition. pp. 5008--5017 (2021)

\bibitem{fang2019data}
Fang, G., Song, J., Shen, C., Wang, X., Chen, D., Song, M.: Data-free
  adversarial distillation. arXiv preprint arXiv:1912.11006  (2019)

\bibitem{1384978}
Fei-Fei, L., Fergus, R., Perona, P.: Learning generative visual models from few
  training examples: An incremental bayesian approach tested on 101 object
  categories. In: 2004 Conference on Computer Vision and Pattern Recognition
  Workshop. pp. 178--178 (2004). \doi{10.1109/CVPR.2004.383}

\bibitem{ge2021yolox}
Ge, Z., Liu, S., Wang, F., Li, Z., Sun, J.: Yolox: Exceeding yolo series in
  2021. arXiv preprint arXiv:2107.08430  (2021)

\bibitem{gou2021knowledge}
Gou, J., Yu, B., Maybank, S.J., Tao, D.: Knowledge distillation: A survey.
  International Journal of Computer Vision  \textbf{129}(6),  1789--1819 (2021)

\bibitem{metric_evaluation}
Gower, J., Legendre, P.: Metric and euclidean properties of dissimilarity
  coefficients. Journal of Classification  \textbf{3},  5--48 (02 1986).
  \doi{10.1007/BF01896809}

\bibitem{he2016deep}
He, K., Zhang, X., Ren, S., Sun, J.: Deep residual learning for image
  recognition. In: Proceedings of the IEEE conference on computer vision and
  pattern recognition. pp. 770--778 (2016)

\bibitem{hinton2015distilling}
Hinton, G., Vinyals, O., Dean, J., et~al.: Distilling the knowledge in a neural
  network. arXiv preprint arXiv:1503.02531  \textbf{2}(7) (2015)

\bibitem{kariyappa2021maze}
Kariyappa, S., Prakash, A., Qureshi, M.K.: Maze: Data-free model stealing
  attack using zeroth-order gradient estimation. In: Proceedings of the
  IEEE/CVF Conference on Computer Vision and Pattern Recognition. pp.
  13814--13823 (2021)

\bibitem{liang2022imitated}
Liang, S., Liu, A., Liang, J., Li, L., Bai, Y., Cao, X.: Imitated detectors:
  Stealing knowledge of black-box object detectors. Association for Computing
  Machinery p. 4839–4847 (2022)

\bibitem{miura2021megex}
Miura, T., Hasegawa, S., Shibahara, T.: Megex: Data-free model extraction
  attack against gradient-based explainable ai. arXiv preprint arXiv:2107.08909
   (2021)

\bibitem{iou_accuracy}
Padilla, R., Netto, S.L., da~Silva, E.A.B.: A survey on performance metrics for
  object-detection algorithms. In: 2020 International Conference on Systems,
  Signals and Image Processing (IWSSIP). pp. 237--242 (2020).
  \doi{10.1109/IWSSIP48289.2020.9145130}

\bibitem{parkhi2012cats}
Parkhi, O.M., Vedaldi, A., Zisserman, A., Jawahar, C.: Cats and dogs. In: 2012
  IEEE conference on computer vision and pattern recognition. pp. 3498--3505.
  IEEE (2012)

\bibitem{yolov3}
Redmon, J., Farhadi, A.: Yolov3: An incremental improvement. arXiv  (2018)

\bibitem{ren2015faster}
Ren, S., He, K., Girshick, R., Sun, J.: Faster r-cnn: Towards real-time object
  detection with region proposal networks. Advances in neural information
  processing systems  \textbf{28} (2015)

\bibitem{romero2014fitnets}
Romero, A., Ballas, N., Kahou, S.E., Chassang, A., Gatta, C., Bengio, Y.:
  Fitnets: Hints for thin deep nets. arXiv preprint arXiv:1412.6550  (2014)

\bibitem{simonyan2014very}
Simonyan, K., Zisserman, A.: Very deep convolutional networks for large-scale
  image recognition. arXiv preprint arXiv:1409.1556  (2014)

\bibitem{szegedy2016rethinking}
Szegedy, C., Vanhoucke, V., Ioffe, S., Shlens, J., Wojna, Z.: Rethinking the
  inception architecture for computer vision. In: Proceedings of the IEEE
  conference on computer vision and pattern recognition. pp. 2818--2826 (2016)

\bibitem{szegedy2013intriguing}
Szegedy, C., Zaremba, W., Sutskever, I., Bruna, J., Erhan, D., Goodfellow, I.,
  Fergus, R.: Intriguing properties of neural networks. arXiv preprint
  arXiv:1312.6199  (2013)

\bibitem{tramer2016stealing}
Tram{\`e}r, F., Zhang, F., Juels, A., Reiter, M.K., Ristenpart, T.: Stealing
  machine learning models via prediction $\{$APIs$\}$. In: 25th USENIX security
  symposium (USENIX Security 16). pp. 601--618 (2016)

\bibitem{truong2021data}
Truong, J.B., Maini, P., Walls, R.J., Papernot, N.: Data-free model extraction.
  In: Proceedings of the IEEE/CVF Conference on Computer Vision and Pattern
  Recognition. pp. 4771--4780 (2021)

\bibitem{xie2017adversarial}
Xie, C., Wang, J., Zhang, Z., Zhou, Y., Xie, L., Yuille, A.: Adversarial
  examples for semantic segmentation and object detection. In: Proceedings of
  the IEEE international conference on computer vision. pp. 1369--1378 (2017)

\bibitem{yan2020cache}
Yan, M., Fletcher, C.W., Torrellas, J.: Cache telepathy: Leveraging shared
  resource attacks to learn $\{$DNN$\}$ architectures. In: 29th USENIX Security
  Symposium (USENIX Security 20). pp. 2003--2020 (2020)

\bibitem{yu2020cloudleak}
Yu, H., Yang, K., Zhang, T., Tsai, Y.Y., Ho, T.Y., Jin, Y.: Cloudleak:
  Large-scale deep learning models stealing through adversarial examples. In:
  NDSS (2020)

\bibitem{zagoruyko2016paying}
Zagoruyko, S., Komodakis, N.: Paying more attention to attention: Improving the
  performance of convolutional neural networks via attention transfer. arXiv
  preprint arXiv:1612.03928  (2016)

\end{thebibliography}

\appendix

\end{document}